\documentclass[%
reprint,
superscriptaddress,
ymb,
prl,
twocolumn,
titlepage,
floatfix,
showpacs,
]{revtex4-2}

\usepackage{graphicx}
\usepackage{hyperref}
\usepackage{amssymb}
\usepackage{amsmath}
\usepackage{textcomp}
\usepackage{xcolor}
\allowdisplaybreaks

\definecolor{linkblue}{RGB}{49,49,148}%prl color
\hypersetup{linkcolor  = linkblue,citecolor  = linkblue,urlcolor   = linkblue,colorlinks = true,}

\makeatletter
\renewcommand*{\eqref}[1]{%
  \hyperref[{#1}]{\textup{\tagform@{\ref*{#1}}}}%
}
\makeatother

\begin{document}
%\title{Disentangling nucleation and domain growth during a laser-induced phase transition in FeRh}
\title{Speed limits of the laser-induced phase transition in FeRh}

\author{M.~Mattern}
\affiliation{Institut f\"ur Physik und Astronomie, Universit\"at Potsdam, 14476 Potsdam, Germany}
\author{J.~Jarecki}
\affiliation{Institut f\"ur Physik und Astronomie, Universit\"at Potsdam, 14476 Potsdam, Germany}
\author{J. A.~Arregi}
\affiliation{CEITEC BUT, Brno University of Technology, 61200 Brno, Czech Republic}
\author{V.~Uhl\'{i}\v{r}}
\affiliation{CEITEC BUT, Brno University of Technology, 61200 Brno, Czech Republic}
\affiliation{Institute of Physical Engineering, Brno University of Technology , 61669 Brno, Czech Republic}
\author{M.~R\"ossle}
\affiliation{Helmholtz-Zentrum Berlin f\"ur Materialien und Energie GmbH, Wilhelm-Conrad-R\"ontgen Campus, BESSY II, 12489 Berlin, Germany}
\author{M.~Bargheer}
\affiliation{Institut f\"ur Physik und Astronomie, Universit\"at Potsdam, 14476 Potsdam, Germany}
\affiliation{Helmholtz-Zentrum Berlin f\"ur Materialien und Energie GmbH, Wilhelm-Conrad-R\"ontgen Campus, BESSY II, 12489 Berlin, Germany}
\email{bargheer@uni-potsdam.de}

\date{\today}

\begin{abstract}
We use ultrafast x-ray diffraction (UXRD) and the polar time-resolved magneto-optical Kerr effect (tr-MOKE) to study the laser-induced metamagnetic phase transition in two FeRh films with thicknesses below and above the optical penetration depth. In the thin film, we identify an intrinsic timescale for the light-induced nucleation of ferromagnetic (FM) domains in the antiferromagnetic material of  $8\,\text{ps}$ that is substantially slower than the speed of sound. For the inhomogeneously excited thicker film, only the optically excited near-surface part transforms within $8\,\text{ps}$. For strong excitations we observe an additional slow rise of the FM phase, which we experimentally relate to a growth of the FM phase into the depth of the layer by comparing the transient magnetization in front- and backside excitation geometry. In the lower lying parts of the film, which are only excited via near-equilibrium heat transport, the FM phase emerges significantly slower than $8\,\text{ps}$ after heating above the transition temperature.
\end{abstract}

\maketitle

First-order phase transitions are characterized by an abrupt change of structural, electronic or/and magnetic properties and a co-existence of multiple phases that introduces nucleation and domain growth to the kinetics of the phase transition \cite{avra1939, gate2017, uhli2016, arre2023, roy2004, jong2013, rand2016, qazi2007, bald2012, keav2018}.

The abrupt change of properties accompanying the emerging phase as a consequence of a fine interplay of spin, charge, orbital and lattice degrees of freedom \cite{jong2013, pole2016} predestine materials featuring first-order phase transitions for ultrafast laser control of functionalities \cite{wegk2014}. In this context, the first-order antiferromagnetic to ferromagnetic (AFM-FM) phase transition of FeRh at $370\,\text{K}$ attracted considerable attention in terms of ultrafast generation of ferromagnetic order \cite{ju2004, berg2006} that extends the more extensively studied ultrafast demagnetization \cite{beau1996, kiri2010} and magnetization reversal \cite{stan2007, radu2011, remy2023}.

The metamagnetic phase transition in FeRh is parameterized by the expansion of the unit cell \cite{arre2020}, the change in the electronic band structure \cite{gray2012, pres2018} and the arising magnetization \cite{maat2005, stam2008}, which each serve as order parameters for different aspects of the ultrafast laser-induced phase transition. Time-resolved photoelectron spectroscopy experiments reveal the formation of an electronic signature of the ferromagnetic state by a photo-induced change of the band structure on a sub-picosecond timescale \cite{pres2021}. X-ray magnetic circular dichroism \cite{unal2017, radu2010}, magneto-optical Kerr effect (MOKE) \cite{mari2012} and double-pulse THz emission spectroscopy \cite{li2022} report a subsequent formation of an in-plane magnetization on a $100\,\text{ps}$ timescale. While the rise of a macroscopic magnetization is dominated by the slow coalescence and alignment of the nucleated domains in an external magnetic field \cite{berg2006,li2022}, the large FM lattice constant as structural order parameter is independent of the orientation of the arising magnetization \cite{mari2012}. Hence, ultrafast x-ray diffraction (UXRD) directly accesses the nucleation and growth of FM domains that also determines the rise of the laser-induced magnetization within the first picoseconds \cite{mari2012, li2022}. Previous UXRD studies report nucleation timescales ranging from $15$ to $90\,\text{ps}$ depending on the probing depth and fluence \cite{mari2012, qui2012}. Thus, the kinetics of the nucleation and growth of the laser-induced FM phase remains unclear and controversial.

Here, we use UXRD experiments on a homogeneously optically excited $12\,\text{nm}$ thin FeRh film to identify an intrinsic fluence-, temperature- and field-independent $8\,\text{ps}$ rise time of the structural order parameter. This nucleation of the FM phase is not limited by the speed of sound, which would require only $2.5\,\text{ps}$. For the inhomogeneously excited $44\,\text{nm}$ thick film, we observe the same intrinsic $8\,\text{ps}$ nucleation timescale and an additional delayed slow rise for high fluences when the deposited energy is sufficient to heat the lower lying parts beyond the transition temperature $T_\text{T}$ by heat diffusion. This unlocks a growth of the FM phase into the depth of the film. Modeling the UXRD results shows that this growth driven by near-equilibrium heat transport is considerably slower than the $8\,\text{ps}$ nucleation timescale after heating above $T_\text{T}$, indicating the crucial role of optically induced non-thermal states for the kinetics of the phase transition. The heat transport timescale is cross-checked by a buried tungsten detection layer, which measures the energy transmitted through FeRh. To complement the insights from the structural order parameter, we probe the subsequent formation of a macroscopic out-of-plane magnetization within $180\,\text{ps}$ by polar tr-MOKE. When we excite the thick FeRh film from the backside, we find a considerable slower and fluence-dependent rise compared to frontside excitation, which experimentally verifies the slow out-of-plane growth of the FM phase.

\begin{figure}[t!]
\centering
\includegraphics[width = \columnwidth]{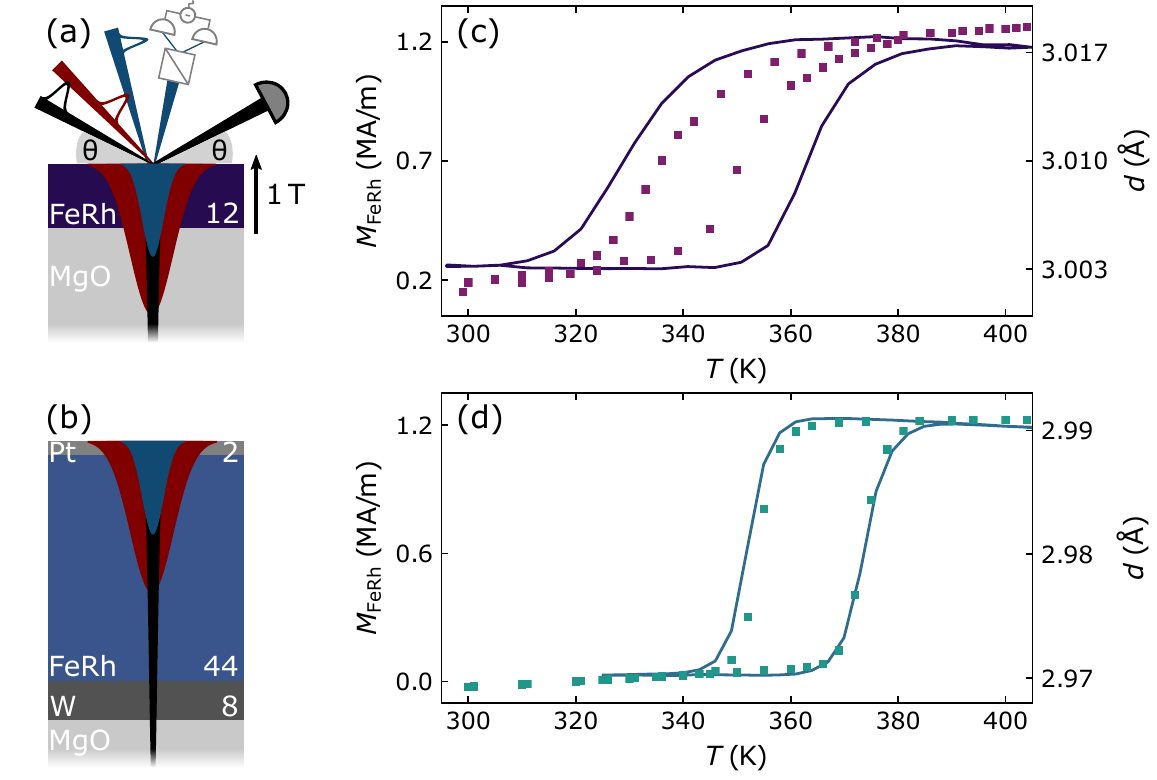}
\caption{\label{fig:fig_1_characterization} \textbf{Characterization of FeRh films:} (a--b) Sketch of the sample structures containing a $12\,\text{nm}$ and a $44\,\text{nm}$ FeRh film, the excitation profile (red) and the probing profile of UXRD (black) and MOKE (blue). The combined UXRD and MOKE experiment is sketched in (a). Panels (c) and (d) compare the temperature-dependent magnetization (solid lines) with the out-of-plane lattice constant (symbols) parametrizing the AFM-FM phase transition in the thin and thick FeRh film, respectively.}
\end{figure}

The two samples are sketched in Fig.~\ref{fig:fig_1_characterization}(a) and (b) and consist of a $12.6\,\text{nm}$ thick FeRh film on MgO(001), and a $43.8\,\text{nm}$ thick FeRh film embedded in a metallic heterostructure on MgO(001) consisting of a $2\,\text{nm}$ Pt capping and a $8\,\text{nm}$ W buffer layer. The thickness of the layers has been characterized via x-ray reflectivity (XRR) and the FeRh films were grown using magnetron sputtering from an equiatomic FeRh target \cite{arre2020}. Both FeRh films exhibit the first-order metamagnetic phase transition characterized by the temperature-dependent magnetization (solid lines) and average out-of-plane lattice constant $d$ (symbols) displayed in Fig.~\ref{fig:fig_1_characterization}(c) and (d). They were measured in thermal equilibrium via vibrating sample magnetometry (VSM) using a QuantumDesign VersaLab magnetometer and XRD performed at the KMC-3 XPP endstation at BESSY II \cite{ross2021}, respectively.

The thin film exhibits a reduced mean transition temperature of $T_\mathrm{T} = 365\,\text{K}$ in comparison to the thick film ($375\,\text{K}$) and a residual FM phase fraction of around $20\,\%$ originating from interface effects \cite{pres2016, fan2010, chen2017}. This reduces the relative expansion associated with the AFM-FM phase transition from $0.6\,\%$ in the thick film to $0.48\,\%$ in the thin film. While the magnetization and lattice constant as order parameters of the temperature-induced phase transition nicely agree for the thick film, the inhomogeneity of the thin film results in a narrower hysteresis for the locally probed lattice constant in contrast to the global magnetization determined by VSM. 

In the combined UXRD and tr-MOKE experiment, the FeRh layers are excited by $p$-polarized pump pulses with a central wavelength of $800\,\text{nm}$ and $100\,\text{fs}$ pulse duration that are incident under $50^\circ$ with respect to the sample surface. Utilizing UXRD, we probe the transient out-of-plane strain response of the FeRh layers via reciprocal space mapping \cite{schi2013a} of the FeRh(002) Bragg peak at a table-top laser-driven plasma x-ray source \cite{schi2012} providing $200\,\text{fs}$ hard x-ray pulses with a photon energy of $\approx 8\,\text{keV}$. The Bragg peak position in reciprocal space is given by the average out-of-plane lattice constant $d$ of the FeRh films via $q_\text{z}=4\pi/d$. Therefore, the laser-induced peak shift accesses its change $\Delta d$ determining the lattice strain $\eta_\text{FeRh}=\Delta d/d_0$ as the relative change with respect to its value $d_0$ before excitation.

\begin{figure}[t!]
\centering
\includegraphics[width = \columnwidth]{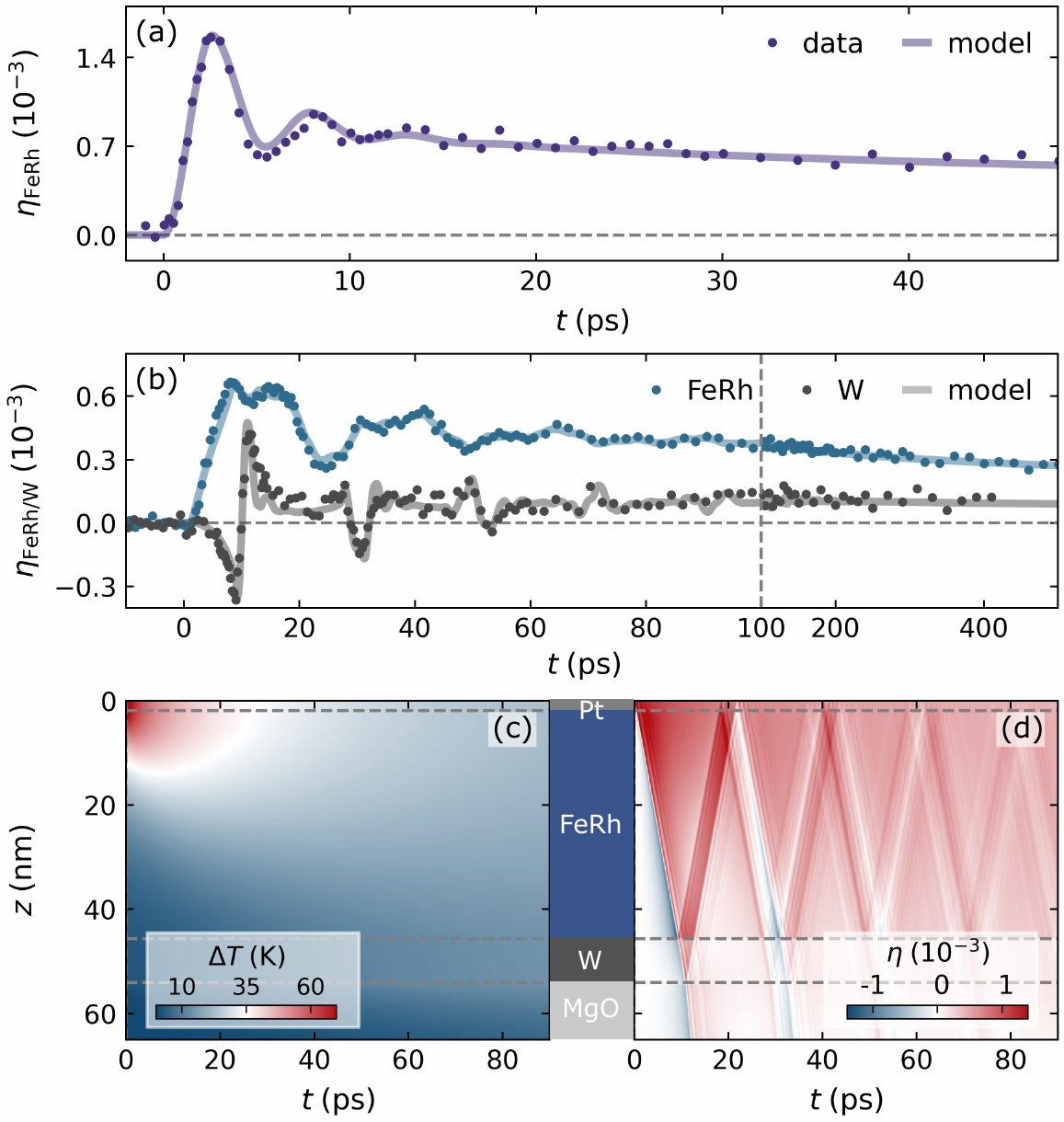}
\caption{\textbf{Thermoelastic strain response calibrated by UXRD:} (a) Transient average strain of the $12\,\text{nm}$ FeRh film (symbols) upon an excitation of $F_\text{st}=0.5\,\text{mJ/cm}^2$. (b) Transient average strain response of the $44\,\text{nm}$ FeRh and the W layer (symbols) for $F_\text{st}=1.5\,\text{mJ/cm}^2$. The solid lines denote our strain model described in the text. The perfect agreement shown in (b) is a strong evidence for the correct modeling of the corresponding the spatio-temporal temperature increase $\Delta T(z,t)$ (c) and strain $\eta(z,t)$ (d).}
\label{fig:fig_4_pux}
\end{figure}

Figure~\ref{fig:fig_4_pux} displays the laser-induced strain response of both FeRh films and the buried W layer to a weak sub-trehsold excitation that is not able to drive the AFM-FM phase transition \cite{mari2012,radu2010}. Thus, the strain response is the superposition of only two contributions: A quasi-static expansion due to heating and coherently driven propagating strain pulses (partially) reflected at the surface and interfaces. In the homogeneously excited $12\,\text{nm}$ FeRh film (optical penetration depth $\lambda=13\,\text{nm}$), the laser excitation launches a strain pulse that is reflected at the surface and partially transmitted into the substrate. This results in a decaying oscillation with a period of $2L_\text{FeRh}/v_\text{s}$ determined by the layer thickness $L_\text{FeRh}$ and the sound velocity $v_\text{s}$ \cite{matt2023} that is superimposed with a decreasing quasi-static expansion due to heat transport into the substrate (see Fig.~\ref{fig:fig_4_pux}(a)). For the thick film sample, the bipolar strain pulse launched at the optically excited surface is partially reflected at the FeRh-W interface, which leads to a more complex shape of the oscillations in the FeRh strain response (see Fig.~\ref{fig:fig_4_pux}(b)). The compression of the W layer is caused by the leading part of the bipolar strain pulse generated in FeRh, and hence essentially senses the profile of optical energy deposition \cite{matt2023}. The W strain turns positive, when the expansive part of the strain pulse enters and the compressive part again exits the layer. The slowly rising expansion of W accesses the heat transport from FeRh into W on tens of picoseconds. These thermoelastic strain contributions scale linearly with the deposited energy \cite{matt2023}. 

In the following, we utilize this calibration of the thermoelastic strain response in the absence of a laser-induced AFM-FM phase transition to extract the transient FM volume fraction $V_\text{FM}$ from the strain response to above-threshold excitations that additionally contains a strain signature from the forming FM phase. For this approach and to access the inhomogeneous spatio-temporal temperature profile within the $44\,\text{nm}$-thick FeRh film, we modeled the strain response by the modular \textsc{Python} library \textsc{udkm1Dsim} \cite{schi2021} utilizing literature values for the thermo-elastic properties stated in Tab.~S1. We calculate the absorption profile and solve the one-dimensional heat diffusion equation determining the spatio-temporal temperature that determines the stress on the lattice \cite{matt2023}. By solving the linear one-dimensional elastic wave equation for this stress we finally calculate the strain response \cite{matt2023} (see supplementary material S1 for a detailed description of the strain modeling). Figures~\ref{fig:fig_4_pux}(c) and (d) display the spatio-temporal strain $\eta$ and temperature increase $\Delta T$ for $F_\text{st}=1.6\,\text{mJ/cm}^2$, respectively. Averaging the strain $\eta(z,t)$ over a respective layer yields the layer-specific strain response measured in our UXRD experiment. We find excellent agreement of the modeled strain response (solid lines) with the measurements in both samples. The excellent agreement with both the $44\,\text{nm}$ FeRh and the buried W layer for a single set of parameters pinpoints the modeled spatio-temporal temperature shown in Fig.~\ref{fig:fig_4_pux}(c). The shape of the initial compression of W quantifies the absorption profile and its slowly rising expansion probes the heat propagated through the inhomogeneously excited FeRh layer \cite{matt2023}.

\begin{figure}[t!]
\centering
\includegraphics[width = \columnwidth]{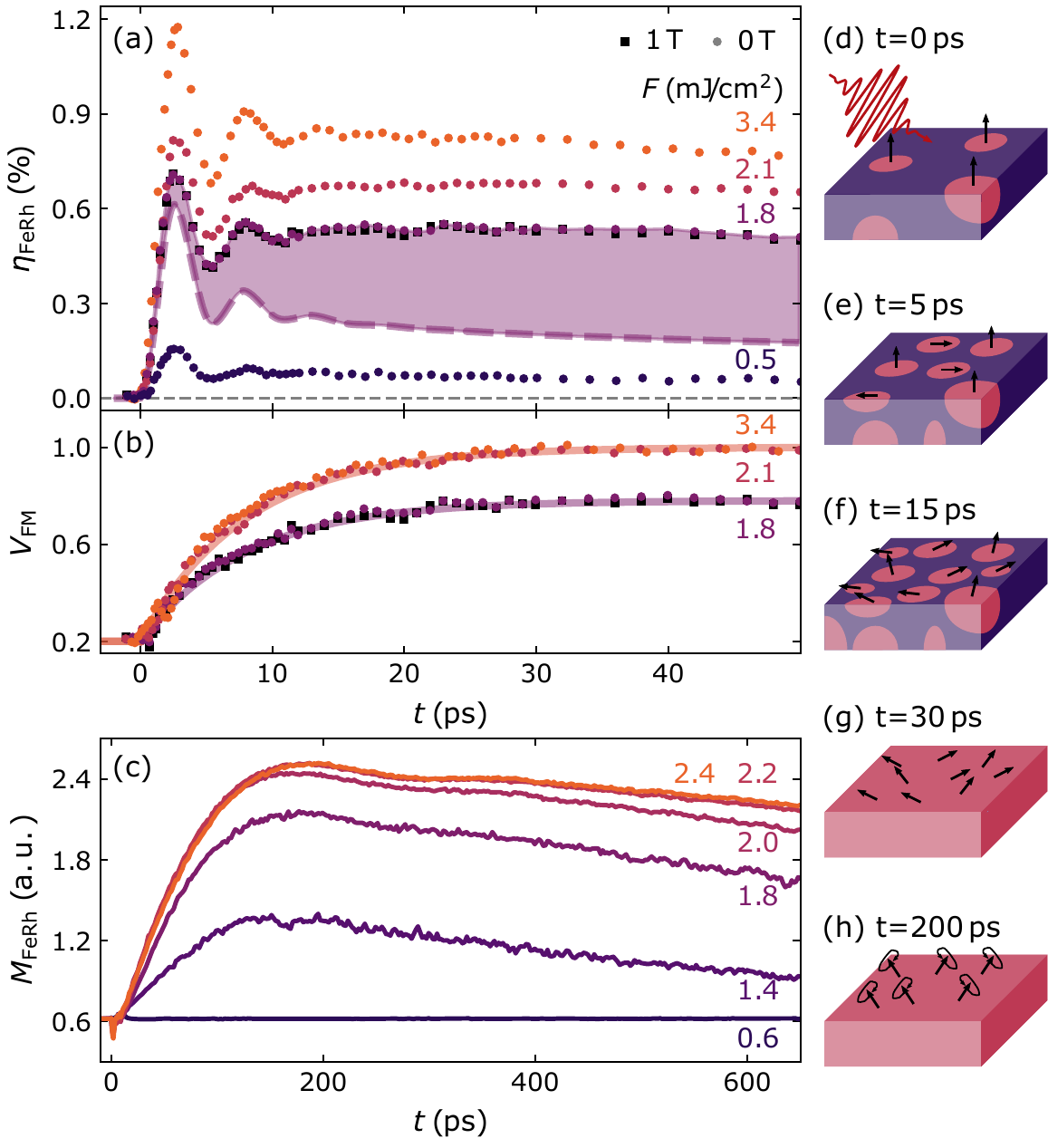}
\caption{\label{fig:fig_2_thin_film} \textbf{Intrinsic timescale of nucleation:} (a) Fluence-dependent strain response of the $12\,\text{nm}$ thick FeRh film. The strain response for a field of $1\,\text{T}$ (squares) matches the one without external field (dots). The dashed line is the modeled strain for $0.5\,\text{mJ/cm}^2$ scaled to $1.8\,\text{mJ/cm}^2$, i.e. the hypothetical response without phase transition.(b) Transient FM volume fraction $V_\text{FM}$ extracted from the difference (coloured area) between measurement and modeled thermoelastic strain. (c) Fluence-dependent transient out-of-plane magnetization from polar MOKE. Field-dependent MOKE data are presented in Fig.~S4 in the supplementary material. Panels (d--h) sketch the kinetics of the phase transition in regard of the structural order parameter (pink color) and the magnetization (arrows) in line with previous results \cite{mari2012,berg2006,li2022, agar2021}. After the nucleation of FM domains the local magnetization precessionally tilts out-of-plane plane (g,h).}
\end{figure}

Figure~\ref{fig:fig_2_thin_film}(a) displays the strain response of the $12\,\text{nm}$-thin FeRh film for various fluences below and above the threshold of the phase transition. The dashed line denotes the modeled thermoelastic strain scaled to $1.8\,\text{mJ/cm}^2$, i.e. the hypothetical strain response without phase transition. The very large difference to the actual measurement (coloured area) highlights the signature of the phase transition in the strain response. The strain measured in an external magnetic field of $1\,\text{T}$ (squares) is identical to the response without an applied magnetic field (dots). This shows that the signature of the phase transition in the structural response is independent of the external magnetic field. We relate the additional strain contribution to $V_\text{FM}(t)$ by considering the expansion of $0.48\,\%$ for a complete phase transition in thermal equilibrium (see Fig.~\ref{fig:fig_1_characterization}(c)). In addition, we considered the thermal expansion coefficient of the FM phase (see Tab.~S1) and that the energy consumed by the latent heat \textcolor{black}{$4.2\,\text{J/gK}$ \cite{rich1973}} at the transition temperature does not contribute to thermoelastic expansion. Figure~\ref{fig:fig_2_thin_film}(b) displays the extracted FM volume fraction $V_\text{FM}(t)$ that parameterizes the laser-driven phase transition. We observe that $V_\text{FM}(t)$ rises as a single exponential according to:
\begin{align}
    V_\text{FM} (t) &= V_\text{FM}^* \cdot \left( 1-e^{-t/\tau} \right) \; ,
\label{eq:eq_2_exponential}
\end{align}
where $V_\text{FM}^*$ denotes the final fraction of the film in the FM phase and $\tau$ is the rise time, independent of the fluence and the applied magnetic field. This model of $V_\text{FM}$ has been discussed and successfully applied in a previous UXRD study of FeRh \cite{mari2012} and assumes the nucleation of FM domains at independent sites according to Avrami \cite{avra1939}. We observe an intrinsic $\tau = 7.8 \pm 0.6\,\text{ps}$ timescale from the fit of all measurements in Fig.~\ref{fig:fig_2_thin_film}(b). This is clearly slower than the relaxation of the lattice: The propagation of strain pulses through the thin film at the speed of sound only takes the time $L_\text{FeRh}/v_s=2.5\,\text{ps}$ (cf. Fig.~\ref{fig:fig_2_thin_film}(a)). We can therefore definitely exclude the widely accepted hypothesis that the sound velocity $v_s$ sets the speed limit for the AFM-FM phase transition in FeRh \cite{mari2012, li2022, kang2023}. Instead, the structural order parameter intrinsically responds on an $8\,\text{ps}$ timescale to direct optical excitation. The transformed volume fraction $V_\text{FM}^*$ increases with fluence and saturates at $F_\text{sat}=2.1\,\text{mJ/cm}^2$, when the complete film is driven into the FM phase. Figure~\ref{fig:fig_2_thin_film}(b) shows that $F=3.4\,\text{mJ/cm}^2$ yields the same $V_\text{FM}^*$, although the thermoelastic strain contributions grow (Fig.~\ref{fig:fig_2_thin_film}(a)).

We complement the insights of the UXRD measurements by measuring the transient out-of-plane magnetization by polar MOKE in the very same experimental setup under identical excitation conditions. We applied a maximum out-of-plane magnetic field of $1\,\text{T}$ provided by an electromagnet. The transient magnetization is extracted by the difference of the transient MOKE signal for opposite field-polarities. The $100\,\text{fs}$-long $p$-polarized probe pulse with a central wavelength of $400\,\text{nm}$ is focussed through the pole of the magnet and is incident under less than $2^\circ$ with respect to the sample normal.

The transient out-of-plane magnetization in Fig.~\ref{fig:fig_2_thin_film}(c) verifies this fluence dependence of $V_\text{FM}^*$. Up to the threshold fluence of $F_\text{th}=0.6\,\text{mJ/cm}^2$ we observe no macroscopic magnetization at all and a fluence of $2.1\,\text{mJ/cm}^2$ fully saturates the laser-induced magnetization. We observe a fluence-independent rise time with the maximum signal at $180\,\text{ps}$ for $1\,\text{T}$, i.e. much slower than the nucleation of the FM domains. This slow formation of a macroscopic magnetization by the alignment of the local magnetization of the nucleated domains was reported previously \cite{mari2012,radu2010,li2022}: The local magnetization initially lies in the sample plane along four equal directions determined by a cubic anisotropy of around $100\,\text{mT}$ that represent the magnetic easy axes due to the shape anisotropy field of $1.38\,\text{T}$ for thin FeRh films \cite{mari2012, cao2008}. For an in-plane magnetic field, a macroscopic in-plane magnetization is formed by field-driven coalescence of the FM domains via domain wall motion in agreement with the observed linear increase of the growth rate of the macroscopic magnetization with the field \cite{li2022}. For the out-of-plane magnetization, in contrast, we observe the maximum to be established faster for smaller out-of-plane fields (Fig.~S4). This is consistent with precessionally tilting \cite{berg2006} the magnetization of the nucleated domains out of the sample plane. Figures~\ref{fig:fig_2_thin_film}(d--h) sketch the series of events.
\begin{figure}[t!]
\centering
\includegraphics[width = \columnwidth]{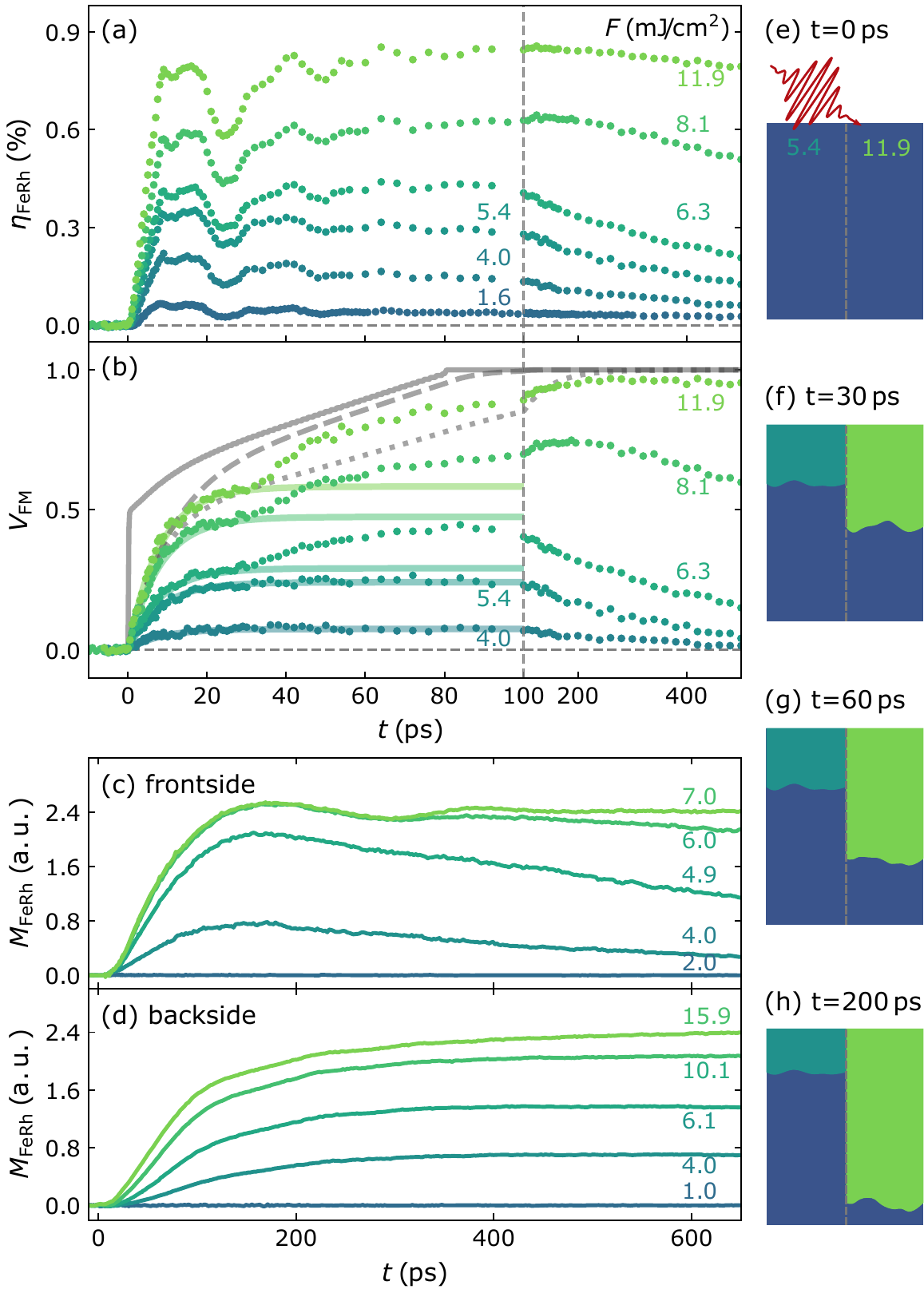}
\caption{\label{fig:fig_3_thick_film} \textbf{Kinetics of out-of-plane growth of the FM phase:} (a) Fluence-dependent field-free strain response of the $44\,\text{nm}$ thick FeRh film. (b) Transient FM volume fraction derived analogously to the thin film by comparing the measured strain to a scaled sub-threshold strain model. (c,d) Fluence-dependent magnetization rise measured by polar MOKE with a magnetic field of $1\,\text{T}$ in front- and backside excitation geometry, respectively. (e--h) Sketch of the out-of-plane growth of the FM phase indicated by the green area for $5.4$ (left) and $11.9\,\text{mJ/cm}^2$ (right) that is induced by a fluence-dependent heating above $T_\text{T}$ at the backside of the film via near-equilibrium heat transport.}
\end{figure}

These MOKE results upon direct optical excitation of the thin film serve as reference for the inhomogeneously excited thick film where the redistribution of energy within the layer by heat transport extends the dynamics of the phase transition. We utilize the finite probing depth of MOKE to gain insights into the out-of-plane evolution of the FM phase by comparing the rising magnetization at the sample surface for front- and backside excitation in Figs.~\ref{fig:fig_3_thick_film}(c and d). For frontside excitation, we observe the same fluence-independent rise time as for the thin film in Fig.~\ref{fig:fig_2_thin_film}(c). In contrast, exciting through the substrate leads to a delayed and slower rise. Additionally, the rise time strongly depends on the fluence (see Fig.~S3 for a comparison normalized to the maximum). The delayed and much slower rise for backside excitation is fluence-dependent, indicating a slow out-of-plane growth of $V_\text{FM}$ via near-equilibrium heat transport that brings the upper part of the film above the transition temperature. However, the tr-MOKE measurements cannot quantify whether it is the heat transport or different kinetics of the phase transition in near-equilibrium that yield the increased timescale.

In order to determine the speed at which the FM phase nucleates and grows in the thick film, we again subtract the fluence-scaled strain response to a sub-threshold excitation $F_\text{st}$ in Fig.~\ref{fig:fig_3_thick_film}(a) measured by UXRD from the strain measured above the threshold. The resulting transient FM volume fraction in Fig.~\ref{fig:fig_3_thick_film}(b) displays the change from a single exponential rise for low fluences to a two-step rise for high fluences ($> 6\,\text{mJ/cm}^2$). For all fluences $V_\text{FM}$ rises according to Eq.~\eqref{eq:eq_2_exponential} with the intrinsic $8\,\text{ps}$ timescale already observed in the thin film during the first $30\,\text{ps}$ (see solid colored lines in Fig.~\ref{fig:fig_3_thick_film}(b)). The amplitude of the additional contribution in the high fluence regime increases with the fluence. The maximum $V_\text{FM}$ is reached at increasing delays up to $250\,\text{ps}$ for the highest fluence, where the complete FeRh film is driven across the AFM-FM phase transition. Reaching the full transformation to the FM phase in Fig.~\ref{fig:fig_3_thick_film}(b) shows that for high fluences the temperature at the backside finally exceeds the transition temperature $T_\text{T}$, thus enabling the phase transition in near-equilibrium. This results in the two-step rise of $V_\text{FM}$ for high fluences in contrast to a phase transition on the $8\,\text{ps}$ timescale limited to the optically excited near-surface region for low fluences as sketched in Fig.~\ref{fig:fig_3_thick_film}(e--h). Note, that the laterally homogeneous growth of the FM phase into the depth, which follows the inhomogeneous optical excitation, differs from the dynamics observed for homogeneous equilibrium heating \cite{gate2017}, where columns through the FeRh film are formed after the nucleation of FM domains at both interfaces before the in-plane domain growth.

To crosscheck our interpretation, we repeated the UXRD varying the initial sample temperature (cf. Fig.~S2). At temperatures only slightly below $T_\text{T}$ the laser excitation is sufficient to drive the phase transition in the not directly excited bottom part of FeRh, which results in a two-step rise of $V_\text{FM}$. At lower temperatures, the heating is insufficient and we observe only a rise on the $8\,\text{ps}$ timescale. In this temperature-dependent experiment, we directly accessed $V_\text{FM}$ by the transient relative amplitudes of the structural Bragg peaks of the AFM and FM phase that are separated in reciprocal space due to the excellent collimation of the x-ray beam at the KMC-3 XPP endstation at BESSY II \cite{ross2021}. This parameter-free analysis also applied by Mariager and co-workers \cite{mari2012} serves as a crosscheck of our findings in Fig.~\ref{fig:fig_2_thin_film} and \ref{fig:fig_3_thick_film}. The slow growth of the FM phase into the depth unlocked by strong excitations explains the different timescales for different fluences and probing depth reported in previous UXRD experiments on inhomogeneously excited FeRh films \cite{mari2012, qui2012}. The exponential probing profile in grazing incidence geometry in previous experiments \cite{mari2012} may have masked the clear two-step behaviour in our experiments.

We obtain additional insights into the growth of the FM phase within the inhomogeneously excited FeRh layer from the modeled spatio-temporal temperature profile $T(z,t)$ in Fig.~\ref{fig:fig_4_pux}(c). From this analysis, we obtain the gray solid line in Fig.~\ref{fig:fig_3_thick_film}(b), which denotes the fraction of the film transiently heated above $T_\text{T}=375\,\text{K}$ for an excitation of $11.9\,\text{mJ/cm}^2$. This analysis shows that already within the first $30\,\text{ps}$ a considerably large fraction and within $80\,\text{ps}$ the complete film is heated above the transition temperature. This is much faster than the observed rise of $V_\text{FM}$ within $250\,\text{ps}$ and indicates that the growth of the FM phase does not simply follow the heating of the backside above $T_\text{T}$ but exhibits intrinsic kinetics. If we assume the FM phase to locally rise with an $8\,\text{ps}$ timescale as soon as the local temperature exceeds $T_\text{T}$, the modeled rise of $V_\text{FM}$ (gray dashed line) is still significantly faster than in the measurement. Therefore, establishing the FM phase by only near-equilibrium heating must be considerably slower than the nucleation in the optically excited near-surface part of the film. The gray dotted line in Fig.~\ref{fig:fig_3_thick_film}(b) represents the combination of an $8\,\text{ps}$ nucleation timescale for the optically heated unit cells and a second $50\,\text{ps}$ timescale for the formation of the FM phase for unit cells heated above $T_\text{T}$ by heat transport. This approach matches the delay when $V_\text{FM}^*$ is reached and provides an estimation of the intrinsic timescale related to the phase transition driven by near-equilibrium heating via thermal electrons and phonons.
%but is not in agreement with the measurement between $30$ and $100\,\text{ps}$. This suggests that the growth into the depth of the film only starts after $30\,\text{ps}$, when the nucleation of the FM domains in the optically excited near-surface-region is finished (see Fig.~\ref{fig:fig_3_thick_film}(b)), even though a large fraction of the layer is already heated above the equilibrium transition temperature.

In summary, we discovered an intrinsic fluence-, temperature- and field-independent $8\,\text{ps}$ timescale for locally establishing macroscopic properties of the FM phase in FeRh via nucleation of domains upon direct optical excitation. This timescale is not limited by the relaxation of the lattice with sound velocity as stated previously \cite{mari2012, kang2023} but represents intrinsic kinetics of the first-order phase transition. In the high fluence regime and for sample temperatures near the transition temperature, we additionally observe a delayed and slow growth of the FM phase into the depth of an inhomogenously excited FeRh layer driven by near-equilibrium heat transport. This observation explains the different rise times of the structural order parameter in previous UXRD experiments \cite{mari2012, qui2012}. Our modeling reveals that the rise of the FM phase after heating above the transition temperature by near-equilibrium heat transport is substantially slower than nucleation driven by direct optical excitation. This hints to the crucial role of modifying the electronic band structure within the first picosecond via photoexcited electrons\cite{pres2021} for the kinetics of the subsequent formation of the equilibrium FM phase.

We acknowledge the DFG for financial support via No.\ BA 2281/11-1 and Project-No.\ 328545488 – TRR 227, project A10. V. U. acknowledges the project CZ.02.01.01/00/22\_008/0004594. Access to the CzechNanoLab Research Infrastructure was supported by the MEYS CR (LM2023051). Beamtimes at the KMC-3 XPP endstation of the synchrotron radiation facility BESSY II at the Helmholtz Zentrum Berlin were required for thorough sample characterization and parameter-free crosscheck measurements.

%\bibliography{references.bib}


\end{document}

\preprint{APS/123-QED}

\title{Supplementary material to: Speed limits of the laser-induced phase transition in FeRh}

\author{M.~Mattern}
\affiliation{Institut f\"ur Physik und Astronomie, Universit\"at Potsdam, 14476 Potsdam, Germany}
\author{J.~Jarecki}
\affiliation{Institut f\"ur Physik und Astronomie, Universit\"at Potsdam, 14476 Potsdam, Germany}
\author{J. A.~Arregi}
\affiliation{CEITEC BUT, Brno University of Technology, 61200 Brno, Czech Republic}
\author{V.~Uhl\'{i}\v{r}}
\affiliation{CEITEC BUT, Brno University of Technology, 61200 Brno, Czech Republic}
\affiliation{Institute of Physical Engineering, Brno University of Technology , 61200 Brno, Czech Republic}
\author{M.~R\"ossle}
\affiliation{Helmholtz-Zentrum Berlin f\"ur Materialien und Energie GmbH, Wilhelm-Conrad-R\"ontgen Campus, BESSY II, 12489 Berlin, Germany}
\author{M.~Bargheer}
\affiliation{Institut f\"ur Physik und Astronomie, Universit\"at Potsdam, 14476 Potsdam, Germany}
\affiliation{Helmholtz-Zentrum Berlin f\"ur Materialien und Energie GmbH, Wilhelm-Conrad-R\"ontgen Campus, BESSY II, 12489 Berlin, Germany}
\email{bargheer@uni-potsdam.de}

\date{\today}
\maketitle

\renewcommand{\thefigure}{S\arabic{figure}}
\renewcommand{\thesection}{S\arabic{section}} 
\renewcommand{\thetable}{S\arabic{table}} 

\section{Modelling the thermoelastic strain response}
In this section, we describe the procedure of modeling the transient strain response by using the modular \textsc{Python} library \textsc{udkm1Dsim} \cite{schi2021} and the layer-specific thermo-elastic parameters given in Tab.~\ref{tab:tab_1_parameter}. In general, the \textsc{udkm1Dsim} library captures the entire series of events in ultrafast x-ray diffraction (UXRD) experiments from the optically deposited energy density to the calculation of a Bragg peak shift via dynamical x-ray diffraction theory.
\begin{table}[b!]
\centering
\begin{ruledtabular}
\begin{tabular}{l c c c c}
 & Pt & FeRh & W & MgO  \\
\hline
unit cell orientation & $(111)$ & $(001)$ & $(001)$ & $(001)$ \\
number of simulated unit cells & & & & \\
$\; \;$ thin film & - & $42 \; (12.6\,\text{nm})$ & - & substrate \\
$\; \;$ thick film & $8 \; (1.8\,\text{nm})$  & $146 \; (43.8\,\text{nm})$ & $28 \; (8.9\,\text{nm})$ & substrate \\
density $\rho \; (\text{g}\,\text{cm}^{-3})$ & $21.45$ & $9.93$ & $19.25$ & $3.58$ \\
elastic constants $(\text{GPa})$ & from \cite{macf1966} & from \cite{palm1975} & from \cite{feat1963} & from \cite{dura1936}\\
$\; \; \; c_{3333}$ & $386$ & $285$ & $523$ & $289.3$ \\
$\; \; \; c_{1133}$ & $232$ & $136$ & $204$ & $87.7$ \\
$\; \; \; c_{2233}$ & $232$ & $136$ & $204$ & $87.7$\\
out-of-plane sound velocity $v_\text{s} \; (\text{nm}\,\text{ps}^{-1})$ & $4.24$ & \underline{$5.00$} $(5.31)$ & $5.21$ & $9.12$ \\
lin. therm. expansion $\alpha \; (10^{-6}\text{K}^{-1})$ & $8.9$ \cite{nix1942} & $9.7$ (AFM) \cite{ibar1994} & $4.6$ \cite{nix1942} & $10.5$ \cite{whit1966} \\
 & & $6.0$ (FM) \cite{ibar1994} & & \\
lin. therm. expansion $\alpha_\perp^\text{uf} \; (10^{-6}\text{K}^{-1})$ & $19.6$ & $19.6$ (AFM) & $8.2$ & $16.9$\\
 & & $12.4$ (FM) & & \\
Gr\"uneisen parameter of electrons $\Gamma_\text{el}$ & $1.4$ \cite{kris2013} & \underline{$1.4$} & \underline{$1.4$} & - \\
Gr\"uneisen parameter of phonons $\Gamma_\text{ph}$ & $2.6$ & $1.7$  & $1.6$ & $1.7$  \\
electron-phonon coupling time $\tau_\text{el-ph} \; (\text{ps})$ & $0.4$ \cite{zahn202} & \underline{$0.6$} & $0.6$ \cite{lin2008} & - \\
heat capacity $C_\text{ph} \; (\text{Jkg}^{-1}\text{K}^{-1})$ & $133$ \cite{shay2016} & $350$ \cite{rich1973} & $132$ \cite{whit1984} & $928$ \cite{barr1959} \\
heat conductivity $\kappa \left( \text{W} \, \text{m}^{-1} \, \text{K}^{-1} \right)$ & & & & \\
$\; \;$ thin film & - & $50$ \cite{berg2006} & - & \underline{$20$} ($50$ \cite{slif1998}) \\
$\; \;$ thick film & $71$ \cite{dugg1970} & \underline{25} ($50$ \cite{berg2006}) & $170$ \cite{chen2019} & \underline{$20$} ($50$ \cite{slif1998}) \\
optical penetration depth $\lambda \; (\text{nm})$ & & & & \\
$\; \;$ thin film & - & multilayer \cite{chen1988} & - & $\inf$ \\
$\; \;$ thick film & \underline{$13$} & \underline{$13$} & \underline{$8$} & $\inf$ \\
\end{tabular}
\end{ruledtabular}
\caption{Thermo-elastic parameters of Pt, FeRh, W and the MgO substrate taken from the literature. Underlined values are optimized in the model to match the observed results. If literature values are available they are provided in parentheses.}
\label{tab:tab_1_parameter}
\end{table}

In our model approach, we do not individually treat electron and phonon degrees of freedom within the framework of a two-temperature model (2TM) since the literature does not provide separate values for the heat conductivity of electrons and phonons, their coupling and the electronic contribution to the thermal expansion in FeRh. Instead, our model only contains a single temperature that corresponds to the phonon temperature after electron-phonon equilibration in a 2TM. The initial temperature profile within the samples is determined by the optical penetration depth $\lambda$ and the macroscopic heat capacity $C_\text{ph}$. The subsequent equilibration of the temperature across the metallic stack and the cooling towards the substrate determines the spatio-temporal temperature increase $\Delta T(z,t)$ by solving the diffusion equation accounting for the layer-specific heat capacities $C_\text{ph}$ and thermal conductivities $\kappa$. The temperature increase determines the laser-induced stress $\sigma^\text{ext}(t)=c_{3333} \cdot \alpha_\perp^\text{uf} \cdot \Delta T(t)$ on ultrafast (uf) timescales via the expansion coefficient $\alpha_\perp^\text{uf}$ that accounts for the exclusive out-of-plane strain response of the metallic films and is calculated by $\alpha_\perp^\text{uf} = (1+\frac{c_{1133}+ c_{2233}}{c_{3333}}) \cdot \alpha$ considering the isotropic expansion coefficient $\alpha$ of bulk specimen \cite{matt2023}. The stress $\sigma^\text{ext}(t)$ drives the spatio-temporal strain response $\eta(z,t)$ as a superposition of a quasi-static expansion and coherently driven propagating strain pulses determined by numerically solving the one-dimensional inhomogeneous elastic wave equation. Finally, the \textsc{udkm1Dsim} library calculates Bragg peaks for each layer considering $\eta(z,t)$ via dynamical x-ray diffraction theory. Their  transient shift is compared to the UXRD measurements.

The values used for the layer-specific thermo-elastic parameters involved in the model process mainly taken from the literature are stated in Tab.~\ref{tab:tab_1_parameter}. Adjusted values are underlined in the table. In addition to the sequence described above, we include a finite stress rise time. In a one-temperature model this corresponds to a phenomenological electron-phonon coupling time $\tau_\text{el-ph}$, which mimics the solution of a 2TM. The different electronic and phononic Gr\"uneisen parameters ($\Gamma_\text{el}$ and $\Gamma_\text{ph}$) parametrize the efficiency of stress generation by energy deposition to the respective degree of freedom. Thus, we re-scale the temperature (energy) increase calculated from the heat diffusion equation by $\left( 1-\frac{\Gamma_\text{ph} - \Gamma_\text{el}}{\Gamma_\text{ph}} \right) \cdot e^{-t/\tau_\text{el-ph}}$ describing the corresponding stress rise. The phononic Gr\"uneisen parameters are calculated from the thermal expansion via $\Gamma_\text{ph}=\alpha_\perp^\text{uf} \frac{c_{3333}}{C_\text{ph}\cdot \rho}$ and the electronic Gr\"uneisen parameters are taken from the literature or optimized to match the shape and amplitude of the propagating strain pulses. The electron-phonon coupling timescale $\tau_\text{el-ph}$ in FeRh is estimated from transient reflectivity measurements. Furthermore, we determine the optical penetration depth of the thick sample by matching the shape of the strain pulse, especially the shape of the rapidly rising compression of the W buffer layer. In contrast, in the thin film we use a transfer-matrix formalism to calculate the absorption in terms of a multilayer model using literature values for the complex refractive index of FeRh \cite{chen1988} and MgO. To match the slow cooling of the optically excited FeRh layer towards W and the MgO substrate in Fig.~2, we reduce the heat conductivity of the FeRh film. We find the same quality of agreement with the data if we take the literature value for the heat conductivity in FeRh but reduce the heat conductivity of the first unit cell of W to $2\,\text{W/Km}$ to model a large interface resistance. Under this assumption, the equilibration of the temperature within the FeRh layer would be even faster. This equally well supports the conclusion in the main text that the domain growth is considerably delayed with respect to the heating above $T_T$.

\section{Temperature-dependent laser-induced AFM-FM phase transition}
In this section, we present the dependence of the laser-induced FM volume fraction on the initial sample temperature of the $44\,\text{nm}$-thick FeRh film that complements the fluence-dependent study in the main text and verifies the approach of extracting $V_\text{FM}$ in the main text by similar findings for a parameter-free more direct approach of extracting $V_\text{FM}$ following \cite{mari2012}.

This experiment was performed at the KMC-3 XPP endstation of BESSY II in the low-alpha operation mode \cite{ross2021} with an x-ray pulse of $17\,\text{ps}$ full-width half-maximum (FWHM) duration. Its parallel and monochromatic x-ray beam in contrast to the convergent beam at the plasma x-ray source \cite{schi2012, schi2013a} enables the separation of the FM and AFM Bragg peaks in reciprocal space. The FM and AFM Bragg peaks represent the increase of the lattice constant across the metamagnetic phase transition. Their integrated intensities serve as a quantitative measure of the respective volume fraction of both phases co-existing during the laser-driven phase transition. The clear separation of the peaks in the synchrotron experiment enables a parameter-free determination of the transient FM volume fraction, and hence a measure of the phase transition kinetics without parametrizing the temperature-dependent thermal expansion as in the main text.
\begin{figure}[t!]
\centering
\includegraphics[width = 0.95\textwidth]{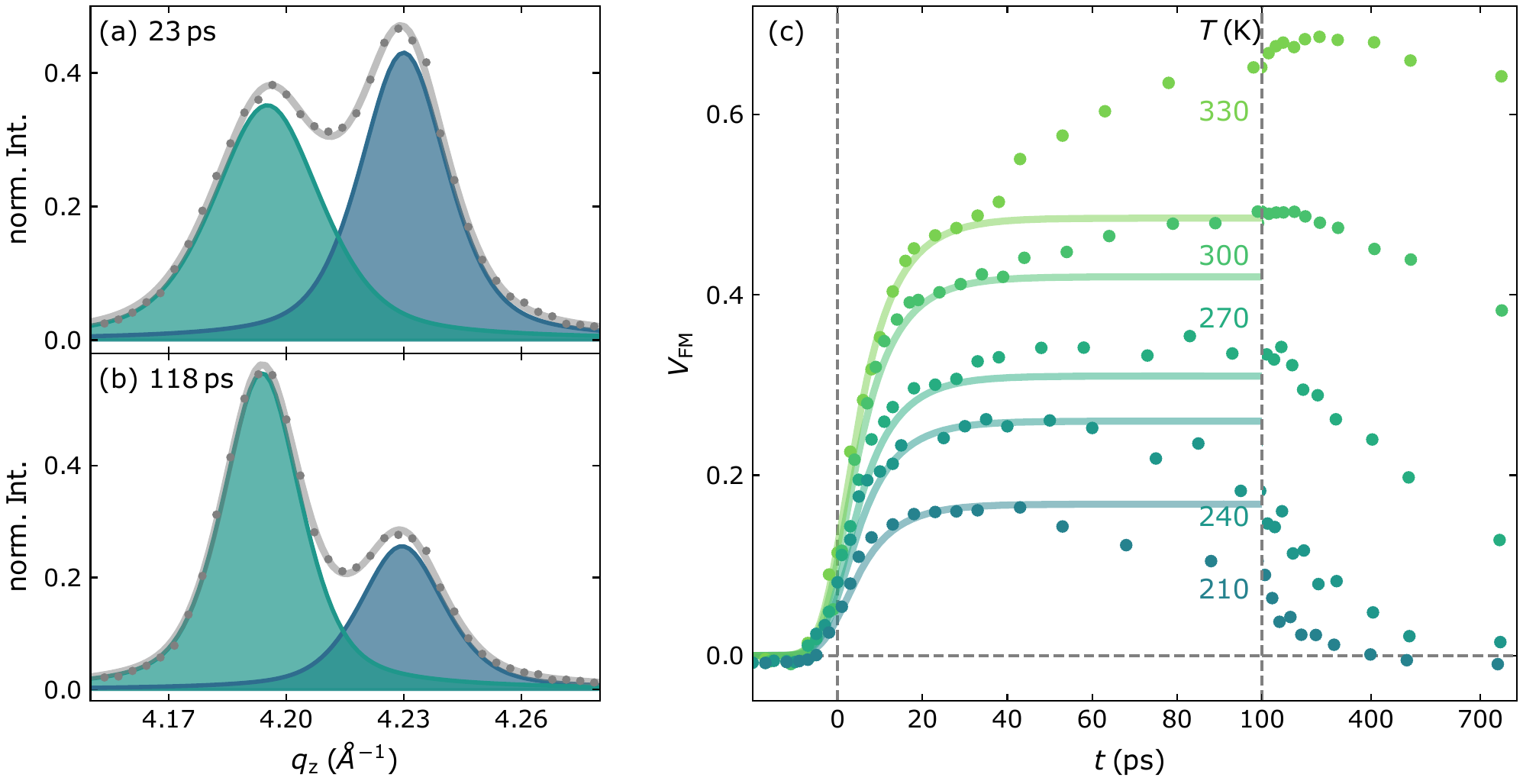}
\caption{\label{fig:fig_S2_temperature_series} \textbf{Two-step rise of FM volume fraction for excitation slightly below the transition temperature:} Transient rocking curve (grey) at $23\,\text{ps}$ (a) and $118\,\text{ps}$ (b) for an excitation of $6.5\,\text{mJ/cm}^2$ at $330\,\text{K}$ separated into the Bragg peaks of the FM (green) and AFM phase (blue). In general, the integral of the Bragg peak is proportional to the fraction of the probed volume exhibiting the respective lattice structure. Therefore, the ratio of the integrated FM and AFM Bragg peak determines the transient FM volume fraction $V_\text{FM}$ following the approach of Mariager and co-workers \cite{mari2012} for different initial sample temperatures in panel (c). The solid lines denote an exponential rise on an $8\,\text{ps}$ timescale convoluted with a Gaussian of $17\,\text{ps}$ FWHM representing the utilized x-ray pulse.}
\end{figure}

Figure~\ref{fig:fig_S2_temperature_series}(a) and (b) display the transient intensity distribution along the reciprocal $q_\text{z}$ coordinate (symbols) encoding the out-of-plane lattice constant $d$ via $q_\text{z}=4\pi/d$ at $23\,\text{ps}$ and $118\,\text{ps}$ after excitation with a $600\,\text{fs}$ laser pulse with a central wavelength of $1028\,\text{nm}$. The modeled intensity distribution (solid grey line) is the superposition of the AFM Bragg peak (blue) and the FM Bragg peak (green) at a larger lattice constant that emerges upon laser-excitation. Finally, the intensity distribution at each pump-probe delay yields $V_\text{FM}$ by the integrated intensity of the FM Bragg peak with respect to the total diffracted intensity \cite{mari2012}. Figure~\ref{fig:fig_S2_temperature_series}(c) displays the transient $V_\text{FM}$ at different initial sample temperatures ranging from $210\,\text{K}$ to $330\,\text{K}$ close to the transition temperature with a time-resolution of $\approx 15\,\text{ps}$ given by the full-width half maximum of the x-ray pulse. We observe a similar behavior with increasing temperature as for increasing fluence in the main text. While the excitation well below the transition temperature only induces nucleation of FM domains on an $8\,\text{ps}$ timescale, increasing the initial sample temperature enables the energy redistribution via heat transport to heat above the transition temperature which causes a growth of the nucleated FM domains into the depth of the film. As for high fluences, this results in a two-step rise of $V_\text{FM}$ that becomes more dominant when approaching the transition temperature. The observation of a two-step rise of the FM volume fraction and a nice match of the first rise by an $8\,\text{ps}$ timescale considering the temporal resolution of the experiment (solid lines in Fig.~\ref{fig:fig_S2_temperature_series}(c)) for the parameter free-extraction of $V_\text{FM}$ following the approach of Mariager and co-workers \cite{mari2012} demonstrates the validity of the approach of extracting $V_\text{FM}$ in the main text.

\cleardoublepage

\section{Fluence-dependent magnetization rise for front- and backside excitation}
\begin{figure}[h!]
\centering
\includegraphics[width = 0.95\textwidth]{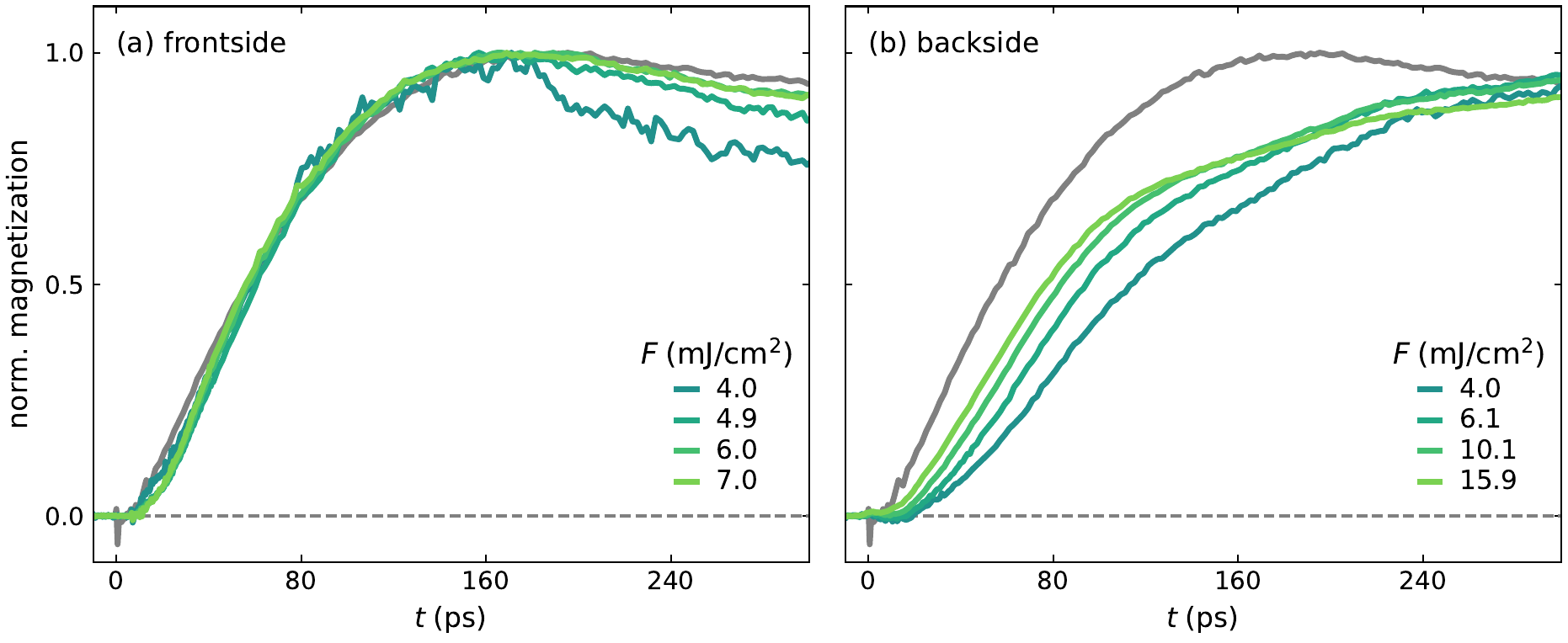}
\caption{\label{fig:fig_S3_norm_moke} \textbf{Fluence-dependent magnetization rise for front- and backside excitation:} Transient magnetization of the thick FeRh film from the main text normalized to the maximum in front- (a) and backside (b) excitation geometry for various fluences. The grey solid line denotes the rise of the normalized magnetization in the thin FeRh film. This representation highlights the fluence-independent rise for frontside excitation in quantitative agreement with the thin film and the decreasing speed of magnetization rise with decreasing fluence for backside excitation that is delayed in respect to the thin film dynamics originating from optically induced nucleation.}
\end{figure}

Figure~\ref{fig:fig_S3_norm_moke} presents the transient magnetization of the thick FeRh film from Fig. 3 in the main text normalized to its maximum. This representation for front- (a) and backside (b) excitation highlights the fluence dependence of the rise of the magnetization indicating the underlying kinetics of the AFM-FM phase transition. As a reference we additionally include the rise of the magnetization of the thin film as solid grey line. In frontside excitation geometry, the rise of the magnetization is independent from the fluence and in quantitative agreement with the behaviour of the thin film indicating optically induced nucleation in the near surface region of the thick FeRh film. In contrast, the rise of the magnetization at the sample surface after backside excitation becomes strongly fluence dependent and slows down with decreasing fluence. In addition, the magnetization rise is considerably slower than in the thin film and slightly delayed even for the highest fluence indicating the intrinsic kinetics of the domain growth driven by near-equilibrium heat transport. With increasing fluence less redistribution of the optically deposited energy density is required to heat the near-surface region above the transition temperature, which speeds up the formation of the FM phase. 

\section{Field-dependent magnetization rise}
\begin{figure}[ht!]
\centering
\includegraphics[width = 0.95\textwidth]{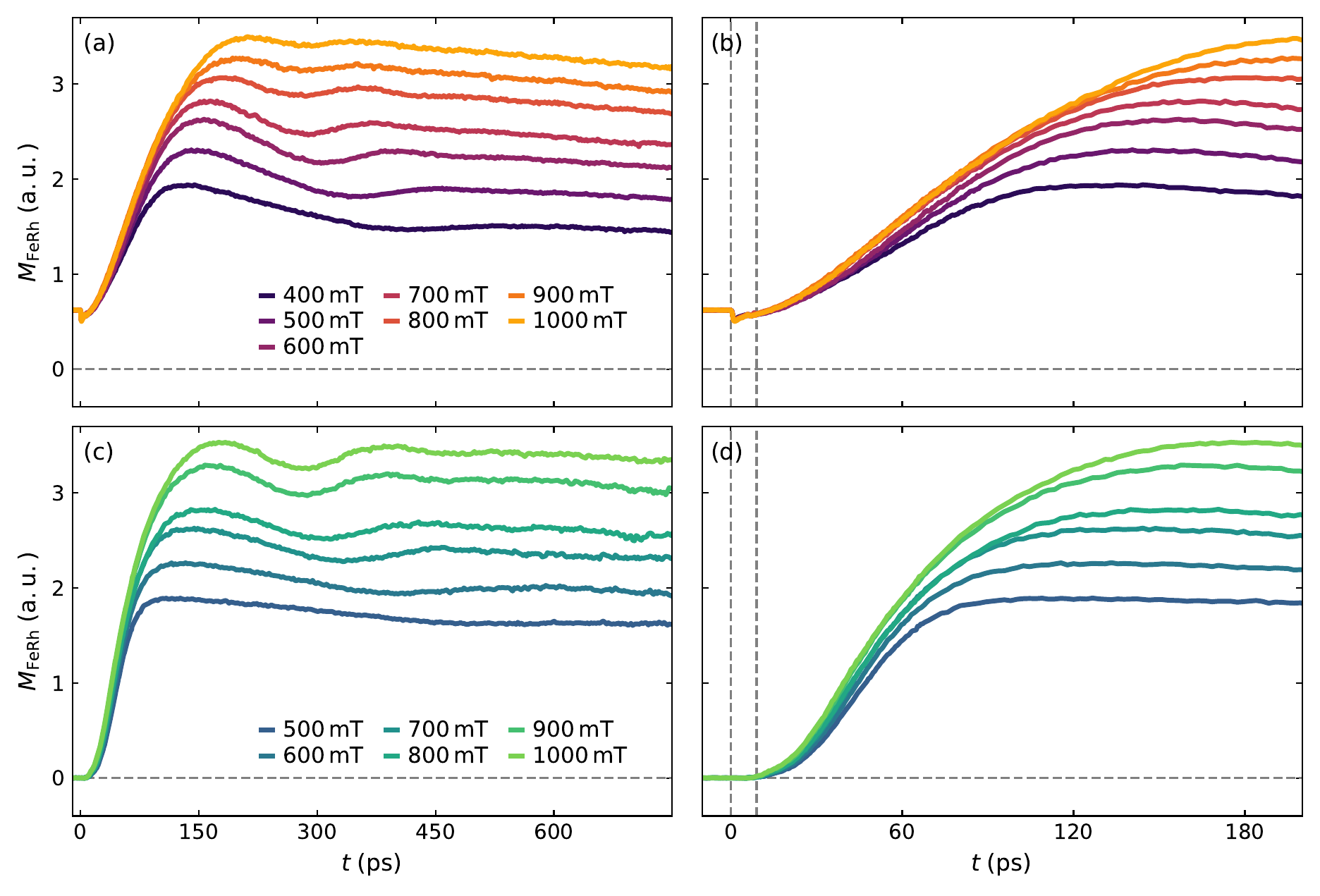}
\caption{\label{fig:fig_S4_field_moke} \textbf{Field-dependent magnetization rise of both FeRh films:} Transient out-of-plane magnetization as function of the external magnetic field strength of the thin (a) and thick FeRh film (c). Panels (b) and (d) display the respective rise of the magnetization within the first $200\,\text{ps}$ upon laser excitation. The vertical grey dashed lines indicate a field-independent latency in the rise of the out-of-plane magnetization of around $10\,\text{ps}$ in agreement with previous work \cite{li2022}. With increasing external magnetic field we observe a faster rise but a later establishing of the maximum out-of-plane magnetization that is not saturated up to $1\,\text{T}$ due to a dominating shape anisotropy of $1.38\,\text{T}$ given by the magnetization of FeRh \cite{cao2008}.}
\end{figure}

Figure~\ref{fig:fig_S4_field_moke} displays the laser-induced out-of-plane magnetization as function of the external magnetic field for both samples. These results extend the discussion in the main text and demonstrate a field-dependent frequency of the heavily damped precession and a field-dependent rise of the magnetization highlighted by the behaviour within the first $200\,\text{ps}$ in panels (b) and (d). Independent from the external magnetic field we observe a latency of the rise of the magnetization of around $10\,\text{ps}$ in agreement with previous work \cite{li2022}. At high magnetic fields the larger maximum out-of-plane magnetization is reached later. The stronger external out-of-plane field tilts the effective field and hence the final direction of the magnetization further out of plane. Therefore, it takes longer to establish the maximum magnetization via the precessional motion of the magnetization. This observation for an out-of-plane magnetic field conceptually differs from the faster rise of an in-plane magnetization with increasing strength of the in-plane magnetic field due to domain wall motion in previous experiments \cite{mari2012,li2022}.

%\bibliography{supplementary_references}